\documentclass[10pt,a4paper]{article}

\usepackage{setspace}
\usepackage{amsmath, latexsym, amssymb}
\numberwithin{equation}{section}
\usepackage[english]{babel}
\usepackage{color}
\definecolor{myred}{rgb}{0.9,0,0}
\definecolor{myblue}{rgb}{0,0,0.75}
\definecolor{mygreen}{rgb}{0,0.8,0}
\definecolor{gray}{rgb}{0.9,0.9,0.9}
\usepackage{ifpdf}

\ifpdf
   \usepackage{graphicx}
    \usepackage[pdftex,
               pdftitle={Analytical VaR and Expected Shortfall in Asymptotic Multi-Factor Credit Portfolio Model.},
               pdfsubject={Risk measurement and management},
               pdfkeywords={analytical VaR, credit portfolio, capital allocation, multi-factor model},
               pdfauthor={
               Mikhail Voropaev
               },
               pdfstartview=FitH,
               bookmarks=false,
               breaklinks=true,
               colorlinks=true,
               citecolor=mygreen,
               linkcolor=myred,
               urlcolor=myblue
               ]{hyperref}
\else
   \usepackage{hyperref}
   \usepackage[dvips]{graphicx}
   \usepackage{rotating}
\fi

\usepackage{listings}
\usepackage[authoryear, round, sort]{natbib}

\begin{document}
\title{Variance-Covariance Based Risk Allocation in Credit Portfolios: Analytical Approximation}
\author{Mikhail Voropaev
\thanks{ING Bank. E-mail:\href{mailto:Mikhail.Voropaev@ingbank.com}{Mikhail.Voropaev@ingbank.com}.\newline
The opinions presented here are those of the author and do not necessarily reflect views of ING Bank.}}
\date{May 2009}
\maketitle

\begin{abstract}
High precision analytical approximation is proposed for variance-covariance based risk allocation
in a portfolio of risky assets. A general case of a single-period multi-factor Merton-type model with
stochastic recovery is considered.
The accuracy of the approximation as well as its speed are compared to and shown to be superior to those of
Monte Carlo simulation.

\end{abstract}


\section{Introduction}
Merton-type models, also referred to as structural models, such as PortfolioManager \citep[][]{PortfolioManager} and CreditMetrics \citep[][]{CreditMetrics}, have become a standard choice for financial institutions' credit risk economic capital frameworks. In these models default correlations between different borrowers are modeled using a set of common systematic risk factors associated with the state of economy. Computationally heavy Monte Carlo simulations are usually used for calculations of portfolio-wide risk measures as well as risk allocation to sub-portfolios and/or individual exposures. However, simulation-based risk allocation on exposure level suffers from Monte Carlo noise and is especially demanding in terms of computer power/time.

Advanced Monte Carlo simulation techniques have been developed in order to improve the stability problem associated with risk decomposition down to individual facilities \citep[see, e.g.,][]{KernelEstimators}. Despite the improvements, the problem exists due to the inherent stochastic nature of Monte Carlo simulations.

Practical demand for stable and fast capital allocation routine in credit portfolios led to development of
analytical techniques. Analytical allocation techniques are also preferable for the purposes of portfolio optimization and risk adjusted pricing. Granularity adjustment \citep[][]{Granularity} and multi-factor adjustment \citep[][]{MultiFactor} are well known extensions of asymptotic single risk factor framework of \citet[][]{Vasicek}.

Unfortunately, most of the research on risk allocation techniques focuses on 'advanced' risk measures like VaR and ES (expected shortfall), leaving variance-covariance based allocation approach aside.
Despite the shortcomings of this approach \citep[see, e.g.,][]{Kalkbrener2004}, this (old-fashioned) allocation method still remains the allocation method of choice for many financial institutions. Variance-covariance based risk contribution is considered to be intuitive, simple and easy to compute risk measure \citep[according to][]{BaselECpractices}. Yet, no efficient analytical solution has been reported so far. A brute force approach consists of calculations of all pairwise correlations in the portfolio (the reader is referred to Section \ref{sec:corr} for the details). The amount of such calculations is quadratic in the number of loans in the portfolio. This quadratic complexity of the calculations make such an approach impractical for big portfolios.

In this article a variance-covariance based analytical risk allocation technique is proposed. The proposed approach is applicable to fully featured Gaussian multi-factor Merton-type models, is suitable for virtually any portfolio size and composition and is remarkably accurate and fast. The main advantage of the proposed technique is that the underlying algorithm is of linear complexity in portfolio size.

It should be pointed out that the author is not an advocate for a particular risk measure. Financial institutions have the responsibility for choosing the risk allocation approach and should clearly  understanding the associated shortcomings. The outcome of any risk allocation technique should be a subject of regular expert reviews and sanity checks.

The technique presented here might be of interest to practitioners building in-house economic capital models. In contrast to most publications focusing on default-only case, the proposed approach is suitable for a very general class of credit portfolio models.

The article is organized as follows. First, a general case of a portfolio of risky assets, whose pairwise correlations are modeled by means of a set of normally distributed common factors, is considered. A routine to allocate portfolio's risk (associated with its standard deviation) to underlying exposures is described. Next, it is demonstrated how to apply the proposed allocation technique to PortfolioManager-like credit portfolio model, i.e. single-period, multi-factor Merton-type model with stochastic recovery and risk neutral valuation at horizon (which accounts for credit migrations). Finally, accuracy and speed of the proposed technique are compared to those of Monte Carlo simulation.


\section{Theory}
In this section a formal theoretical framework for variance-covariance based risk allocation is described. First, it is shown how the Euler allocation principle is applied to portfolio's standard deviation and a link with risk allocation in credit portfolio is briefly described. Next, Gaussian multi-factor model is considered and a series expansion for pairwise covariance is proposed. Finally, an algorithm for calculation of the risk contributions is described.
\subsection{Standard deviation and risk allocation}\label{sec:ulc}
Standard deviation is a subadditive risk measure allowing Euler allocation \citep[see, e.g.,][]{Euler} of portfolio's risk to sub-portfolios and/or individual exposures. Consider a portfolio consisting of risky assets with standard deviations $\{\sigma_i\}$ and pairwise value correlations $\{\rho_{ij}\}$. The standard deviation of the portfolio $\sigma_p$,
\begin{equation}\label{eq:std}
\sigma_p = \sqrt{\sum_i\sigma_i^2 + \sum_{i\neq j}\sigma_i\sigma_j\rho_{ij}},
\end{equation}
can be written as a sum of Euler risk measures (risk contributions) $\sigma_i^c$ as
\begin{equation}
\sigma_p = \sum_i\sigma_i^c=\sum_i\sigma_i\frac{\partial\sigma_p}{\partial \sigma_i}.
\end{equation}
The risk contributions $\{\sigma_i^c\}$ can readily be calculated by differentiation and can be expressed in terms of covariance (or correlation $\rho_{ip}$) between individual exposures and the portfolio
\begin{equation}\label{eq:ulc}
\sigma_i^c = \sigma_i\rho_{ip} = \frac{\mathrm{cov}(i,p)}{\sigma_p} = \frac{1}{\sigma_p}\sum_j\text{cov}(i,j).
\end{equation}

The variance-covariance approach to capital allocation in credit portfolios utilizes the above decomposition to distribute the \emph{economic capital} of the portfolio between individual exposures proportionally to the risk weights \eqref{eq:ulc} as follows
\begin{equation}\label{eq:ec}
(\emph{capital charge})_i = \frac{\sigma_i^c}{\sigma_p}\times (\emph{total economic capital}).
\end{equation}

\subsection{Variance-covariance structure of Gaussian multi-factor framework}\label{sec:corr}
Let us start by considering a portfolio of risky instruments $\{v_i\}$.
The value $v_i$ of each instrument is assumed to be a function of normally distributed random variable $\epsilon_i$. Correlations between variables $\{\epsilon_i\}$ are modeled through a set of normally distributed independent variables $\{\eta_k\}$ referred to as common factors.
Each variable $\epsilon_i$ is split in a sum of instrument specific part, which depends on a Gaussian variable $\xi_i$, and systematic part, which depends on the common factors, as follows
\begin{equation}\label{eq:split}
\epsilon_i = r_i~\sum_k(\beta_i)_k\eta_k + \sqrt{1-r_i^2}~\xi_i.
\end{equation}
The independently distributed random variables $\{\{\xi_i\},\{\eta_k\}\}$\footnote{Assuming all these variables to be independently distributed is equivalent to an assumption that each borrower in the portfolio is represented by one facility. This assumption will be relaxed in Section \ref{sec:application}} are assumed to have zero mean and unit variance. Instrument specific constants $r_i$ and $\{(\beta_i)_k\}$ determine dependency of $\epsilon_i$ on the common factors.
The so-called factor loadings $\{(\beta_i)_k\}$ are subject to normalization condition
\begin{equation}\label{eq:betanorm}
\sum_k(\beta_i)_k^2 = 1.
\end{equation}

 Using the representation \eqref{eq:split}, one can readily calculate a covariance $\rho_{ij}$ between any pair of distinct variables $\{\epsilon_i\}$
\begin{equation}\label{eq:assetcovar}
\rho_{ij} = \left<\epsilon_i \epsilon_j\right> =
r_ir_j\sum_k(\beta_i)_k(\beta_j)_k = r_ir_j\vec{\beta_i}\vec{\beta_j},\quad i\neq j.
\end{equation}
The value covariance between two distinct instruments $\left<v_iv_j\right>$ can be written as
\begin{eqnarray}\label{eq:valcovar}
\left<v_iv_j\right> = \int \! v_i(\epsilon_i)v_j(\epsilon_j)n_2(\epsilon_i,\epsilon_j,\rho_{ij})\mathrm{d}\epsilon_i\mathrm{d}\epsilon_j -
\int\! v_i(\epsilon_i)n(\epsilon_i)\mathrm{d}\epsilon_i \int\! v_j(\epsilon_j)n(\epsilon_j)\mathrm{d}\epsilon_j, \qquad i\neq j,
\end{eqnarray}
where $n$ and $n_2$ are normal and bivariate normal density functions defined as
\begin{eqnarray}
n(x) = \frac{1}{\sqrt{2\pi}}e^{-x^2/2}, \quad
n_2(x,y,\rho) = \frac{1}{2\pi\sqrt{1-\rho^2}}\mathrm{exp}\left(-\frac{x^2-2\rho xy +y^2}{2(1-\rho^2)}\right).
\end{eqnarray}

In theory, the above expression and \eqref{eq:ulc} are sufficient to calculate the standard deviation based risk contributions $\{\sigma_i^c\}$. In practice, however, one is facing a problem of computing $N(N-1)/2$ two-dimensional integrals in \eqref{eq:valcovar} for a portfolio of $N$ instruments. Such a brute force approach
becomes inappropriate for large portfolios, i.e. $N=10^4$ or higher.

\subsection{Series expansion for covariance}
 Aiming for a linear (in $N$) complexity algorithm for computing the risk contributions \eqref{eq:ulc}, let us reduce the dimensionality of the integral in \eqref{eq:valcovar}. Applying Mehler's formula \citep[for the proof see, e.g.,][]{Mehler}
\begin{equation}
\sum_{n=0}^\infty\mathrm{He}_n(x)\mathrm{He}_n(y)\frac{\rho^n}{n!} = \frac{1}{\sqrt{1-\rho^2}} \mathrm{exp}\left(\frac{2\rho xy - \rho^2(x^2+y^2)}{2(1-\rho^2)}\right)
\end{equation}
to the bivariate normal density function, the following expression is obtained for the pairwise value covariance
\begin{equation}\label{eq:covhermite}
\left<v_iv_j\right> = \sum_{n=1}^{\infty}\frac{\rho_{ij}^n}{n!}\int v_i(\epsilon_i)\mathrm{He}_n(\epsilon_i)n(\epsilon_i)\mathrm{d}\epsilon_i  \int v_j(\epsilon_j)\mathrm{He}_n(\epsilon_j)n(\epsilon_j)\mathrm{d}\epsilon_j,
\end{equation}
where $\text{He}_n(x) = (-1)^ne^{x^2/2}(d/dx)^ne^{-x^2/2}$ are Hermite polynomials \citep[for definition and properties of Hermite polinomials see, e.g.,][]{Abramowitz}.
The last expression can be rewritten as
\begin{equation}\label{eq:covarseries}
\left<v_iv_j\right> = \sum_{n=1}^{\infty}\rho_{ij}^nv_i^{(n)}v_j^{(n)}, \quad v^{(n)} = \frac{1}{\sqrt{n!}}\int v(\epsilon)\mathrm{He}_n(\epsilon)n(\epsilon)\mathrm{d}\epsilon.
\end{equation}

The above series converges at least as fast as a geometric series with quotient $\rho_{ij}$. To see this, one can use inequality \citep[see, e.g.,][]{Abramowitz}
\begin{equation}
\text{He}_n(x) < k\sqrt{n!}e^{x^2/4}, \quad k \approx 1.086435,
\end{equation}
which leads to
\begin{equation}
|v^{(n)}| \leq \frac{1}{\sqrt{n!}}\int |v(\epsilon)||\mathrm{He}_n(\epsilon)|n(\epsilon)\mathrm{d}\epsilon < \frac{k}{\sqrt{2\pi}}\int |v(\epsilon)|e^{-\epsilon^2/4}\mathrm{d}\epsilon.
\end{equation}
The last integral in the above inequality is finite for any reasonable value function $v(\epsilon)$.

\subsection{Linear complexity algorithm}\label{sec:algorithm}
Using the series expansion \eqref{eq:covarseries} and expression \eqref{eq:assetcovar}, a linear complexity algorithm can be constructed for calculation of risk contributions \eqref{eq:ulc} as follows.

First, the parameters $v_i^{(n)}$, $n=1\ldots n_{max}$ are evaluated for each instrument $i=1\ldots N$. The number of terms $n_{max}$ used for calculations determines an accuracy of the algorithm. Second, assuming $N_f$ common factors, the following portfolio specific parameters are calculated
\begin{equation}\label{eq:portparam}
P^{(n)}_{k_1\ldots k_n} = \sum_{i=1}^Nr^n_iv^{(n)}_i(\beta_i)_{k_1}\ldots(\beta_i)_{k_n}, \quad k_j = 1\ldots N_f.
\end{equation}
Third, the risk weights \eqref{eq:ulc} are calculated using the following expression
\begin{equation}\label{eq:covarcontrib}
\sigma_p\sigma_i^c = \left<v_i^2\right> + \sum_{j, j\neq i}\left<v_iv_j\right>,
\end{equation}
which after substituting \eqref{eq:covarseries}, \eqref{eq:assetcovar}  and \eqref{eq:portparam} becomes
\begin{equation}\label{eq:approxulc}
\sigma_p\sigma_i^c = \sigma_i^2 + \sum_{n=1}^{n_{max}}r^n_iv^{(n)}_i\sum_{k_1\ldots k_n}^{N_f}(\beta_i)_{k_1}\ldots(\beta_i)_{k_n}P^{(n)}_{k_1\ldots k_n}  - \sum_{n=1}^{n_{max}}\left(r^n_iv^{(n)}_i\right)^2.
\end{equation}
Finally, the standard deviation of the portfolio and the risk charges per instrument are calculated as
\begin{equation}
\sigma_p = \sqrt{\sum_i\sigma_p\sigma_i^c}, \quad \sigma_i^c = \frac{\sigma_p\sigma_i^c}{\sqrt{\sum_i\sigma_p\sigma_i^c}}.
\end{equation}

Under certain circumstances the algorithm may be further simplified. Consider, for example, a single factor model. Expressions \eqref{eq:portparam} and \eqref{eq:approxulc} then become
\begin{equation}
\sigma_p\sigma_i^c = \sigma_i^2 + \sum_{n=1}^{n_{max}}r^n_iv^{(n)}_iP^{(n)}  - \sum_{n=1}^{n_{max}}\left(r^n_iv^{(n)}_i\right)^2, \quad P^{(n)} = \sum_{i=1}^Nr^n_iv^{(n)}_i.
\end{equation}
Another simplification takes place if we consider a default-only credit portfolio, i.e.
\begin{equation}
v(\epsilon) = \left\{
\begin{array}{rl}
1 & \text{if } \epsilon > d \\
1 - l & \text{if } \epsilon \leq d
\end{array}
\right.
\end{equation}
In this case the parameters $v^{(n)}$ can be computed analytically
\begin{equation}\label{eq:defonly}
v^{(n)} = l\frac{e^{-d^2/2}}{\sqrt{2\pi n!}}\text{He}_{n-1}(d)
\end{equation}

\section{Practice}\label{sec:practice}
In order to assess the accuracy of the proposed risk allocation routine, PortfolioManager-like credit portfolio model is considered. In this section basic description of the model is given. Some aspects concerning application of the proposed risk allocation framework to the credit portfolio model are discussed. The analytically calculated risk measures are compared to those obtained as a result of Monte Carlo simulation. It is finally concluded that the accuracy of the analytical approximation is higher than the accuracy of Monte Carlo simulation with $10^8$ scenarios.


\subsection{Credit portfolio model}\label{sec:portmodel}
Consider Gaussian multi-factor Merton-type credit portfolio model with stochastic recovery rate. The credit portfolio consists of loans whose values $\{v_i\}$ are functions of the corresponding borrowers' asset returns $\{\epsilon_i\}$. The asset returns are assumed to be normally distributed with the correlation structure as described in Section \ref{sec:corr}.

A borrower is considered to be in default at some point in time if its asset return falls below the borrower-specific threshold, i.e. when $\epsilon < -d$. The quantity $d$ is called \emph{distance to default} and is related to the borrower's \emph{probability of default} $p$ through the standard normal cumulative distribution function $\Phi$ as
\begin{equation}\label{eq:pd}
p = \Phi(-d) = \int_{-\infty}^{-d}\frac{e^{-\epsilon^2/2}}{\sqrt{2\pi}}\mathrm{d}\epsilon.
\end{equation}

The portfolio is observed at some time in the future $t_h$ called \emph{horizon}. Assume that both probability of default till horizon $p$ and (cumulative) probability of default till maturity $p_m$ are known for each loan in the portfolio. In case a loan matures before the horizon, its probability of default at horizon is assumed to be equal to its probability of default at maturity.

In case of default, the value is defined by the recovery rate $(1-l)$ and loan's risk-free value $v_0e^{-r(t_m-t_h)}$ as
\begin{equation}\label{eq:defaultval}
v(\epsilon) = (1-l)v_0e^{-r(t_m-t_h)}, \qquad \epsilon \leq -d,
\end{equation}
where $r$ is a risk-free rate and $t_m$ is loan's maturity. In case of no default and maturity before or at horizon, the risk-free value of the loan is recovered
\begin{equation}\label{eq:riskfreeval}
v(\epsilon) = v_0e^{-r(t_m-t_h)}, \quad t_m \leq t_h, \qquad \epsilon > -d.
\end{equation}
Risk neutral valuation, as described in \citet[][]{Vasicek}, is applied in no default case to the loans maturing after horizon
\begin{equation}\label{eq:nodefaultval}
v(\epsilon) = v_0e^{-r(t_m-t_h)}\left(1-l\cdot\Phi\left(b\sqrt{\frac{t_m}{t_m-t_h}}-\epsilon\sqrt{\frac{t_h}{t_m-t_h}}\right)\right), \quad t_m > t_h, \qquad \epsilon > -d,
\end{equation}
where
\begin{equation}
b = \Phi^{-1}(p_m) + \lambda r\frac{t_m-t_h}{\sqrt{t_m}}.
\end{equation}
The $\lambda$ in the above expression stands for the so-called \emph{market price of risk} and $r$ is the borrower-specific parameter introduced in \eqref{eq:split}. The above valuation takes into account change of loan's market value due to \emph{credit migration}.

Uncertainty in recovery rates in case of default is modeled by means of Beta distribution \citep[see, e.g.,][]{Abramowitz} whose probability density function is
\begin{equation}\label{eq:beta}
f(x,a,b) = \frac{1}{B(a,b)}x^{a-1}(1-x)^{b-1},
\end{equation}
where $a$ and $b$ are parameters characterizing the distribution. The above parameters are chosen to math the mean $l$ and the variance $\sigma^2$ of the loss distribution
\begin{equation}\label{eq:betameanvar}
l = \frac{a}{a+b}, \quad \sigma^2 = \frac{l(1-l)}{a+b+1} = \frac{l(1-l)}{k},
\end{equation}
where $k$ is a parameter defining the shape of the distribution. Losses in case of default are assumed to be distributed independently from asset returns.

\subsection{Applying risk allocation routine}\label{sec:application}
In order to adopt the risk allocation procedure proposed in Section \ref{sec:algorithm}, some modifications should be made to \eqref{eq:covarcontrib} and \eqref{eq:approxulc}.

 For the value function $v(\epsilon)$ of a loan at horizon which is defined by \eqref{eq:defaultval}-\eqref{eq:nodefaultval}, the standalone variance $\sigma^2$ has to be calculated as
\begin{equation}
\sigma^2 = \int v^2(\epsilon)n(\epsilon)\mathrm{d}\epsilon - \left(\int v(\epsilon)n(\epsilon)\mathrm{d}\epsilon\right)^2
+ p\frac{l(1-l)}{k}\left(v_0e^{-r(t_m-t_h)}\right)^2,
\end{equation}
where the last term takes into account loss uncertainty in case of default according to \eqref{eq:betameanvar}. Since distribution of losses in case of default is assumed to be independent, the other terms in \eqref{eq:approxulc} are not affected by stochastic nature of recovery rate.

Expressions \eqref{eq:covarcontrib} and \eqref{eq:approxulc} were derived assuming that all the borrower-specific parts of asset returns ($\{\xi_i\}$ in \eqref{eq:split}) are independent. In practice, however, some borrowers are represented by multiple loans in the portfolio. Such loans have perfectly correlated asset returns, which makes the series expansion \eqref{eq:covarseries} invalid. It is still possible to modify the expressions \eqref{eq:covarcontrib} and \eqref{eq:approxulc}, preserving the linear complexity of the risk allocation algorithm.

Consider a borrower $a$ with multiple loans $\{v_{ia}\}$. Expression \eqref{eq:covarcontrib} has to be modified as follows
\begin{equation}
\sigma_p\sigma_i^c = \sum_{j}\left<v_{ia}v_{ja}\right> + \sum_{j,b\neq a}\left<v_{ia}v_{jb}\right>.
\end{equation}
Introducing $V_a(\epsilon_a) = \sum_iv_{ia}(\epsilon_a)$, net value of the loans of the same borrower, the first term in the above expression can be written as $\left<v_{ia}V_{a}\right>$. Assuming the recovery distributions of the loans of the same borrower perfectly correlated, the following modified version of \eqref{eq:approxulc} can be derived
\begin{equation}\label{eq:approxulccredit}
\sigma_p\sigma_i^c = \left<v_{ia}V_{a}\right> + \sum_{n=1}^{n_{max}}r^n_iv^{(n)}_{ia}\sum_{k_1\ldots k_n}^{N_f}(\beta_i)_{k_1}\ldots(\beta_i)_{k_n}P^{(n)}_{k_1\ldots k_n}  - \sum_{n=1}^{n_{max}}r^{2n}_iv^{(n)}_{ia}V^{(n)}_a,
\end{equation}
where the first term is defined as
\begin{equation}\label{eq:approxulccredit2}
\left<vV\right> = \int v(\epsilon)V(\epsilon)n(\epsilon)\mathrm{d}\epsilon - \int v(\epsilon)n(\epsilon)\mathrm{d}\epsilon \int V(\epsilon)n(\epsilon)\mathrm{d}\epsilon
+ p\frac{l(1-l)}{k}v_0V_0e^{-2r(t_m-t_h)}.
\end{equation}

\subsection{Numerical results}
\begin{figure}[t]
\centering
\ifpdf
$\begin{array}{ccc}
\includegraphics[width=0.35\textwidth,viewport=90 480 390 740,clip]{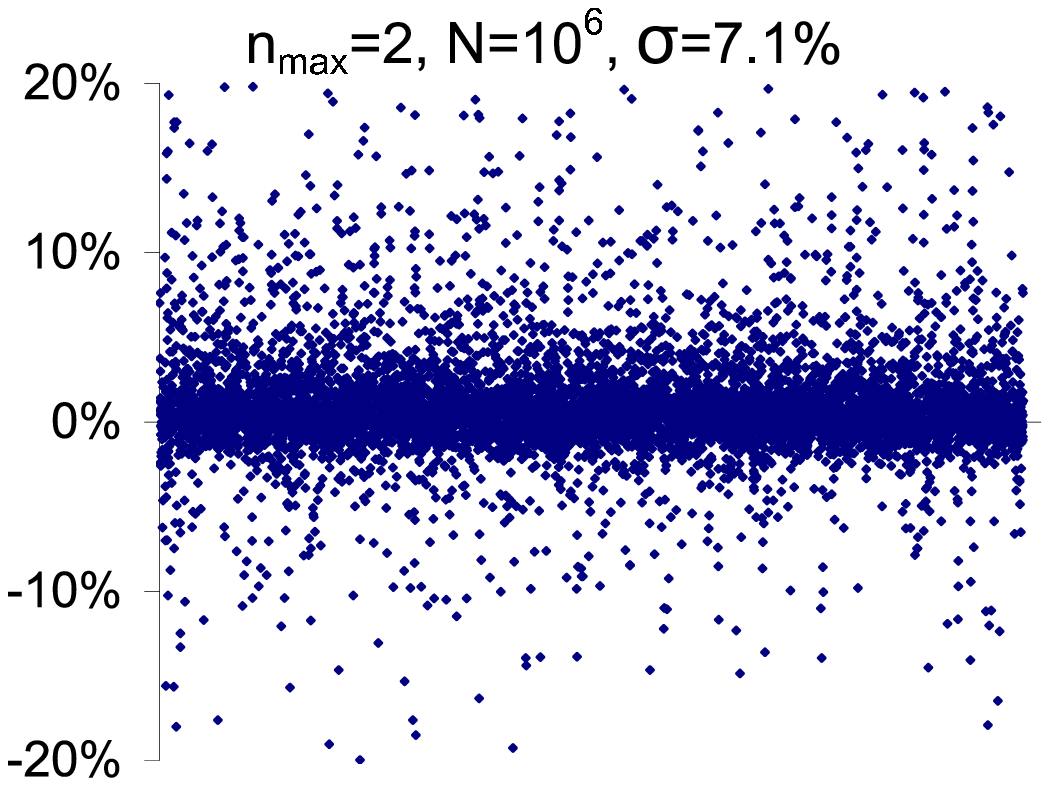} &
\includegraphics[width=0.35\textwidth,viewport=90 480 390 740,clip]{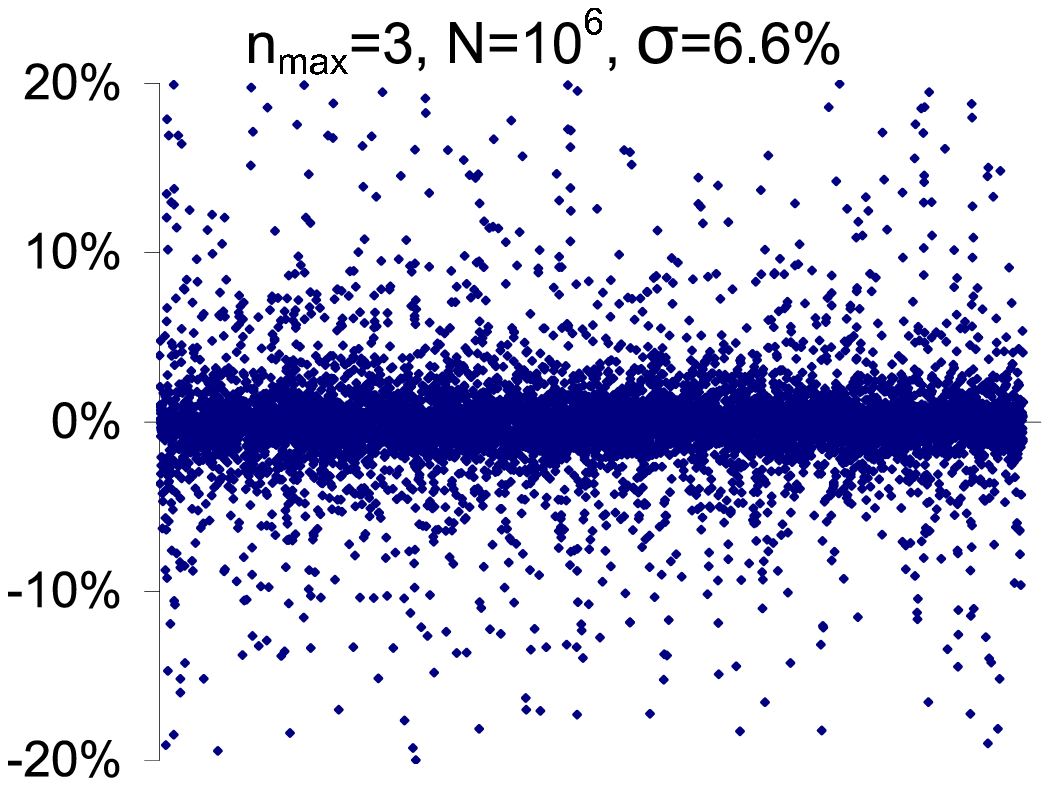} \\
\includegraphics[width=0.35\textwidth,viewport=90 480 390 740,clip]{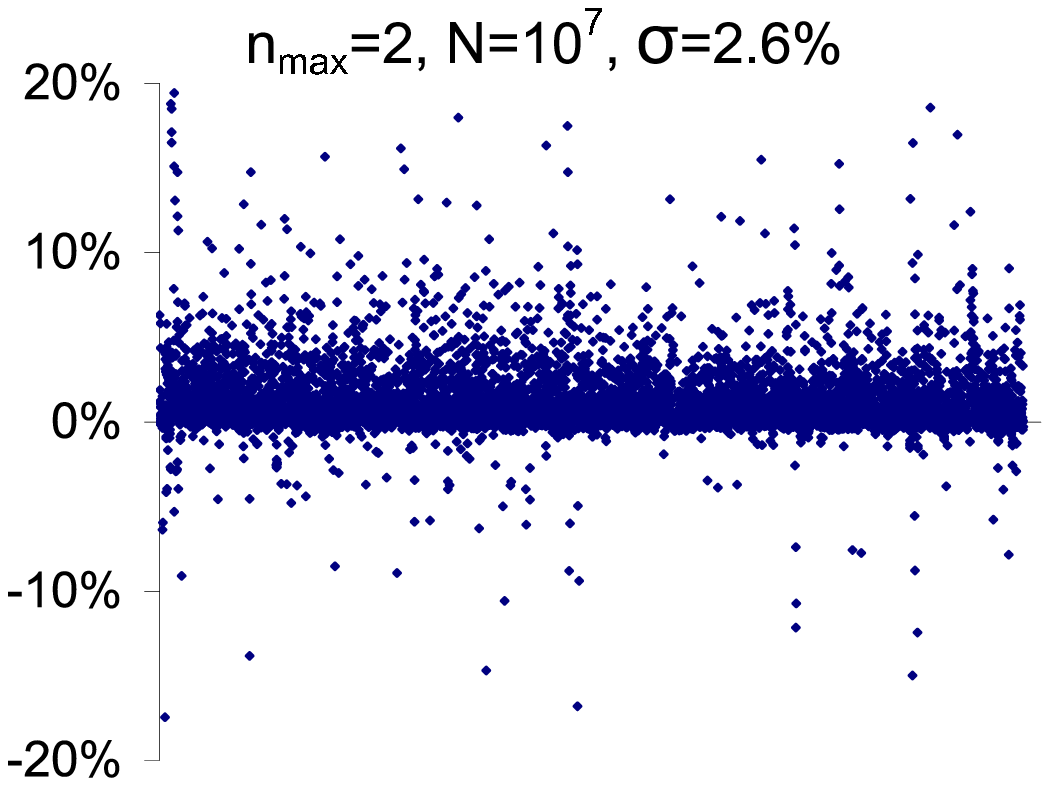} &
\includegraphics[width=0.35\textwidth,viewport=90 480 390 740,clip]{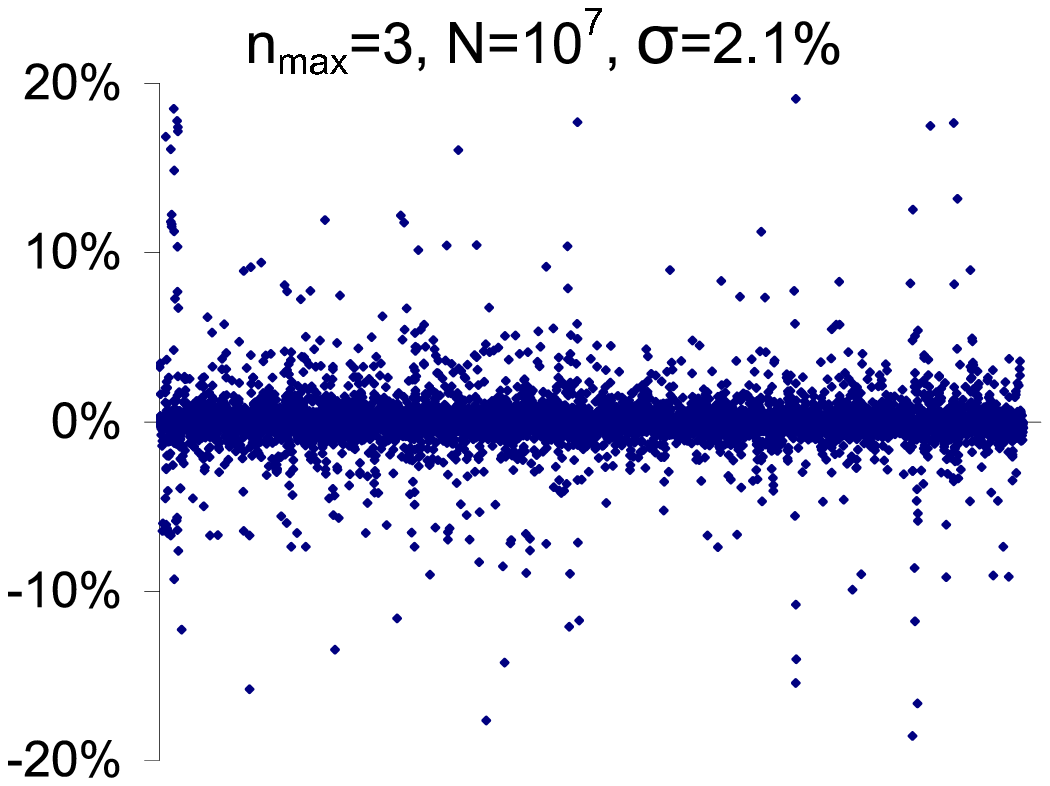} \\
\includegraphics[width=0.35\textwidth,viewport=90 480 390 740,clip]{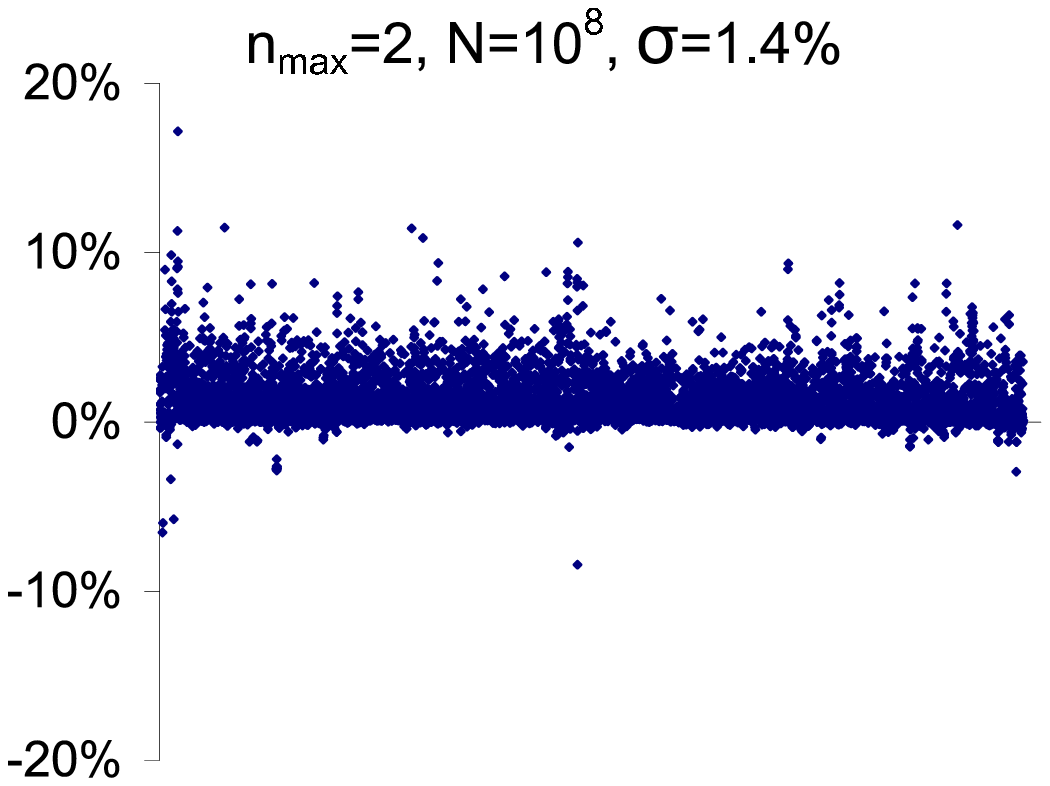} &
\includegraphics[width=0.35\textwidth,viewport=90 480 390 740,clip]{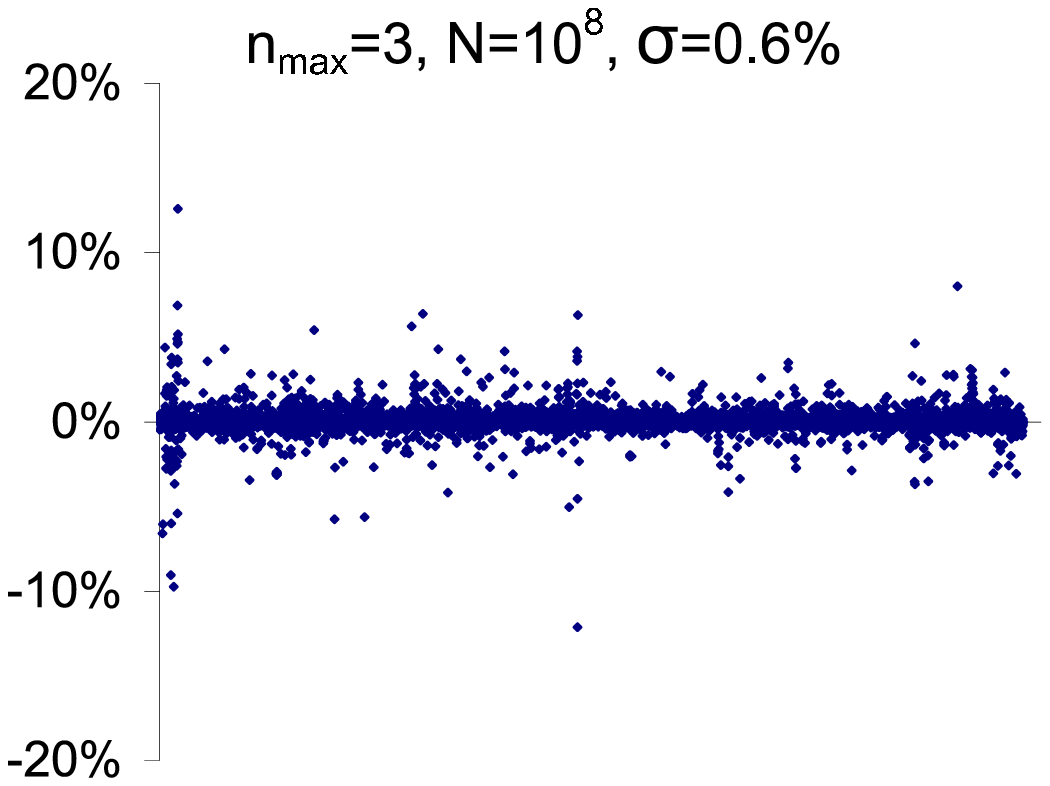}
\end{array}$
\else
$\begin{array}{ccc}
\includegraphics[width=0.35\textwidth,viewport=0 0 300 260,clip]{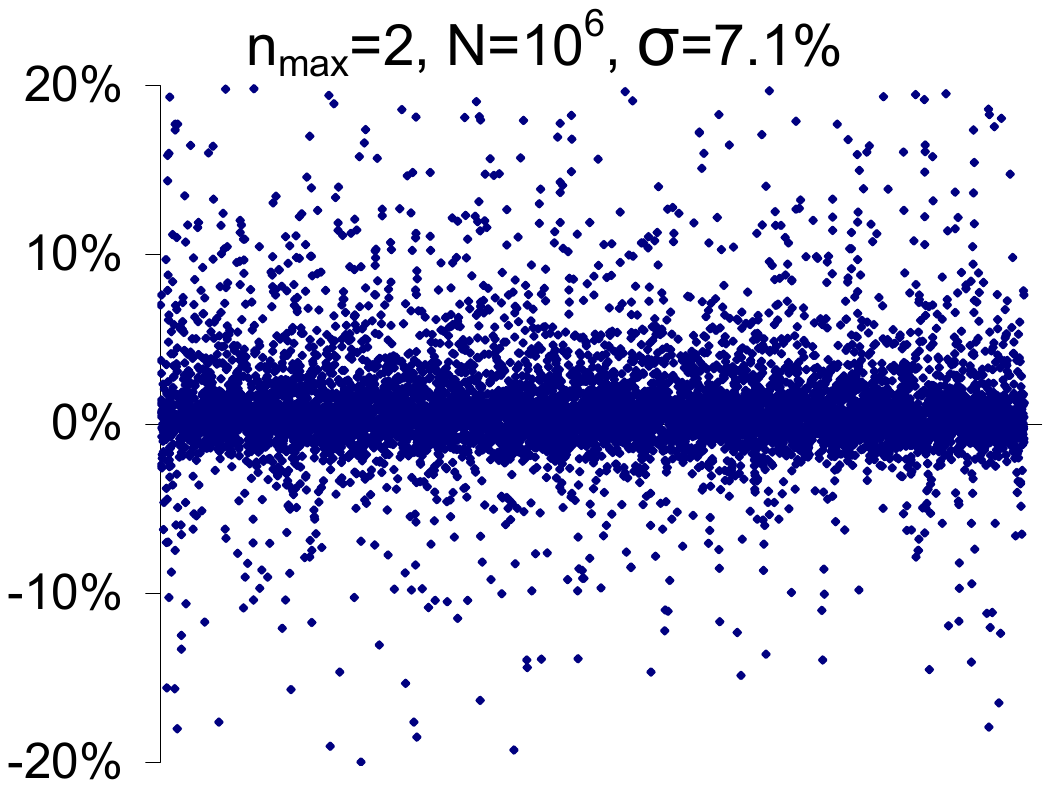} &
\includegraphics[width=0.35\textwidth,viewport=0 0 300 260,clip]{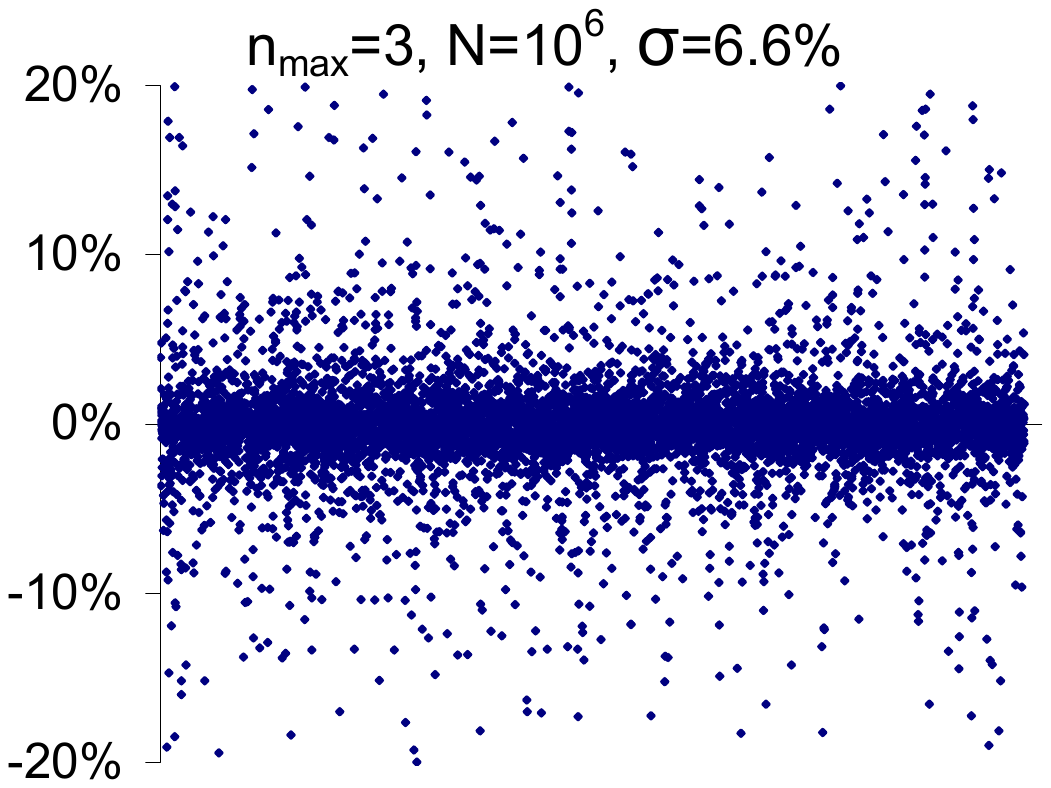} \\
\includegraphics[width=0.35\textwidth,viewport=0 0 300 260,clip]{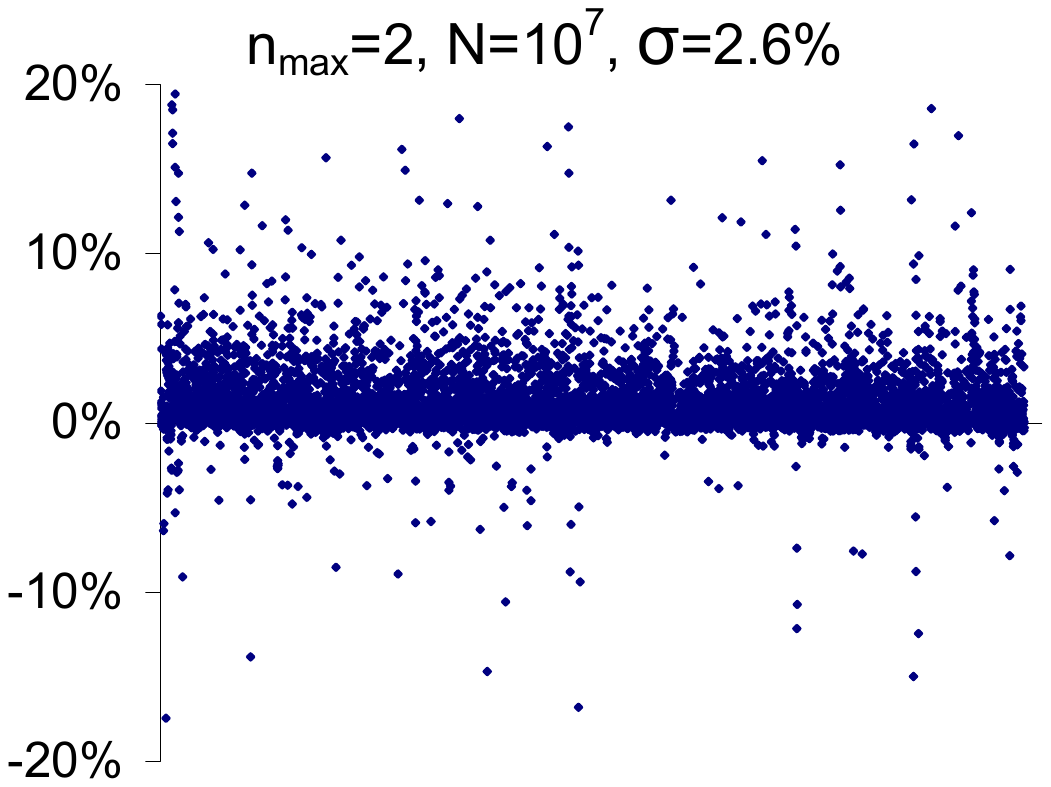} &
\includegraphics[width=0.35\textwidth,viewport=0 0 300 260,clip]{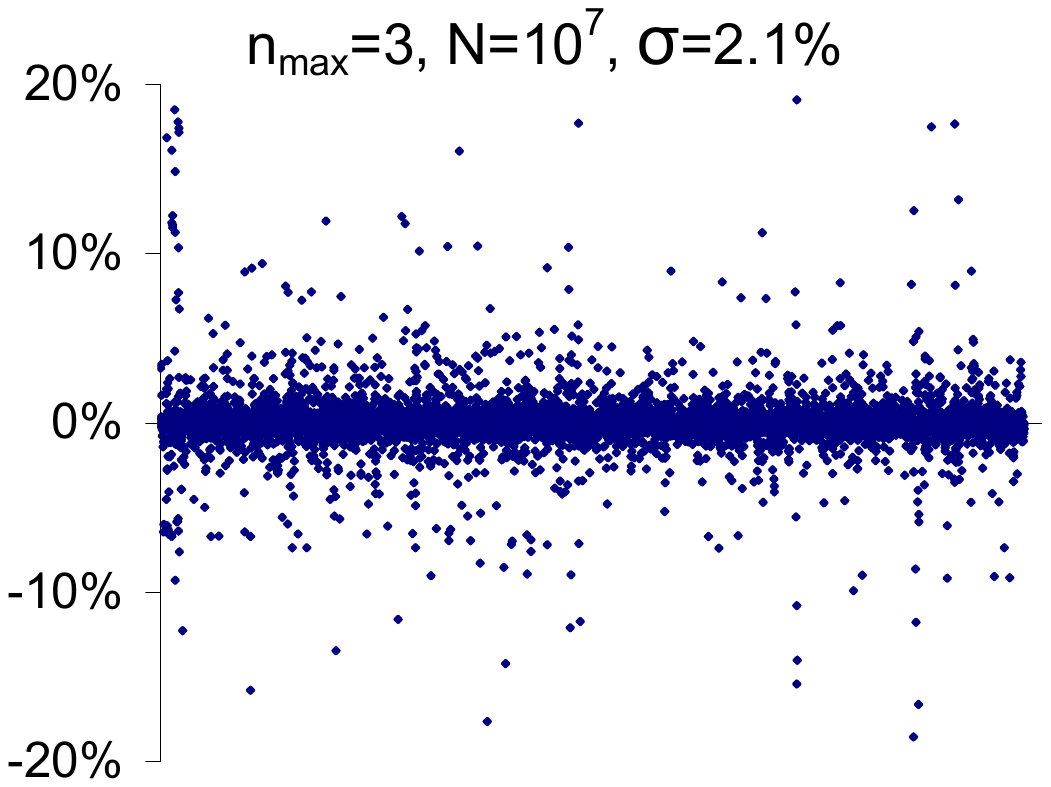} \\
\includegraphics[width=0.35\textwidth,viewport=0 0 300 260,clip]{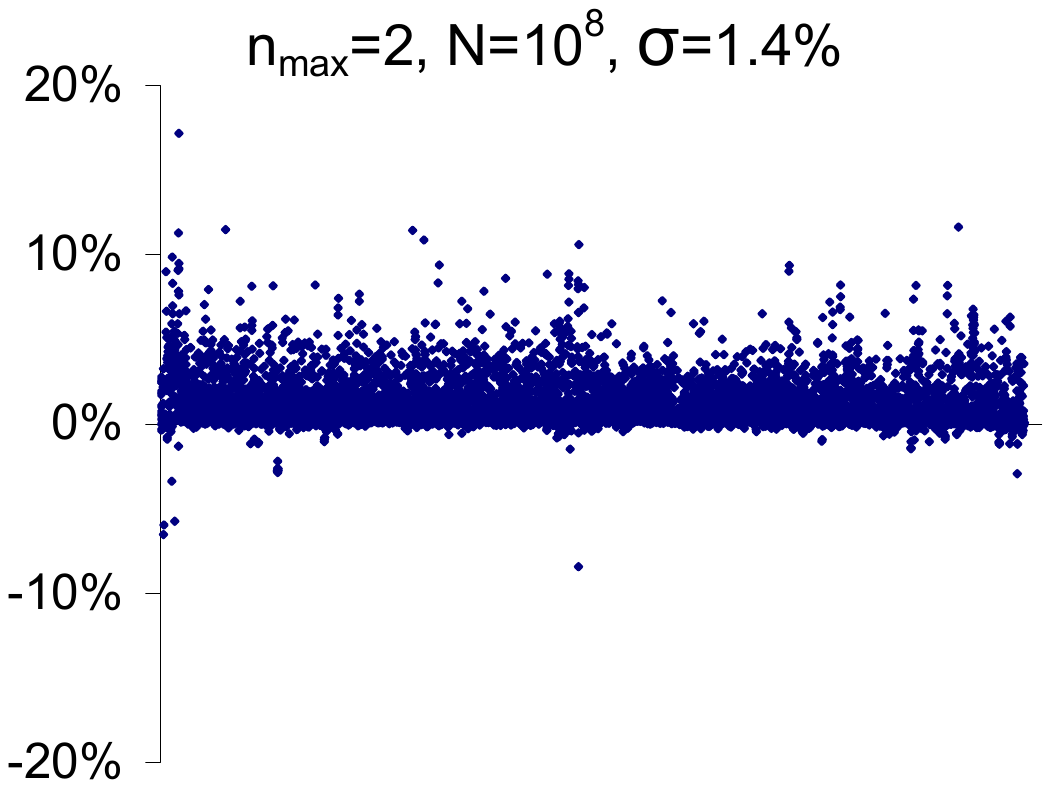} &
\includegraphics[width=0.35\textwidth,viewport=0 0 300 260,clip]{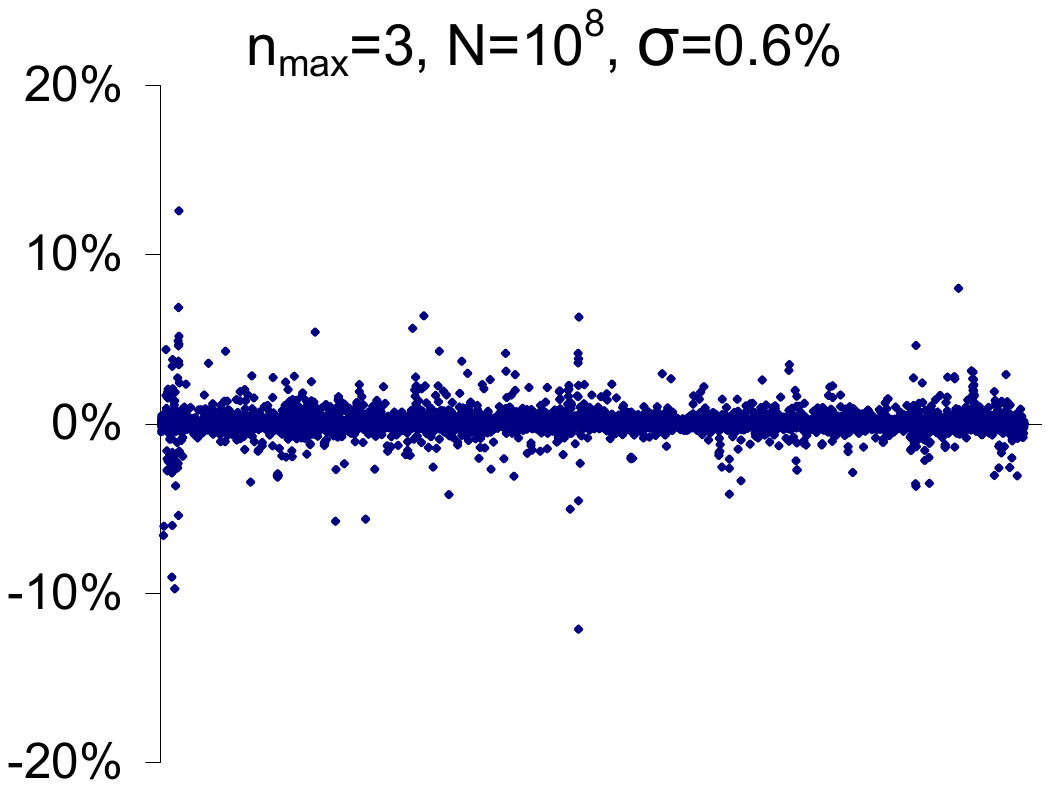}
\end{array}$
\fi
\parbox{0.7\textwidth}{
\caption[Relative differences between Monte Carlo and analytical estimates of the risk contributions.]
{\emph{Relative differences between Monte Carlo and analytical estimates of the risk contributions $\sigma^c$.}}\label{fig:error}}
\end{figure}

\begin{figure}[t]
\centering
\ifpdf
$\begin{array}{ccc}
\includegraphics[width=0.4\textwidth,viewport=60 550 310 720,clip]{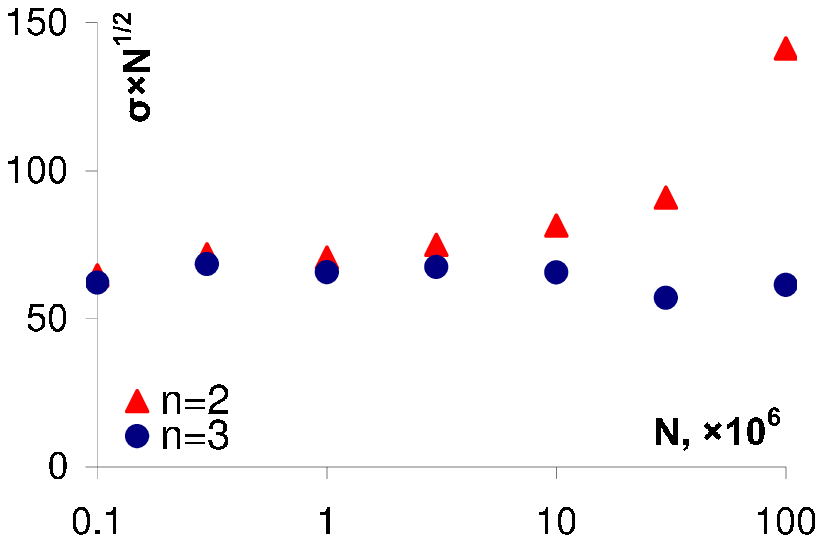} &
\includegraphics[width=0.35\textwidth,viewport=50 520 310 720,clip]{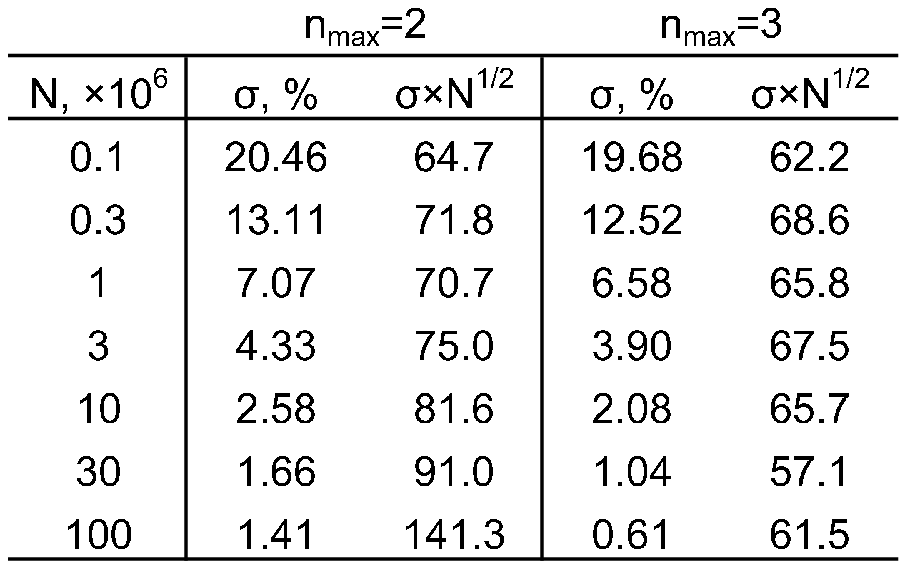}
\end{array}$
\else
$\begin{array}{ccc}
\includegraphics[width=0.4\textwidth,viewport=0 0 250 170,clip]{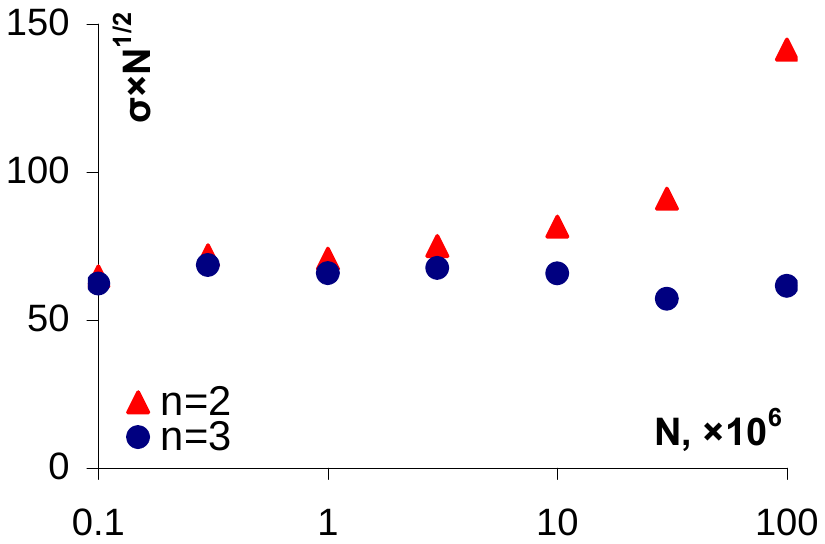} &
\includegraphics[width=0.35\textwidth,viewport=-10 -30 250 170,clip]{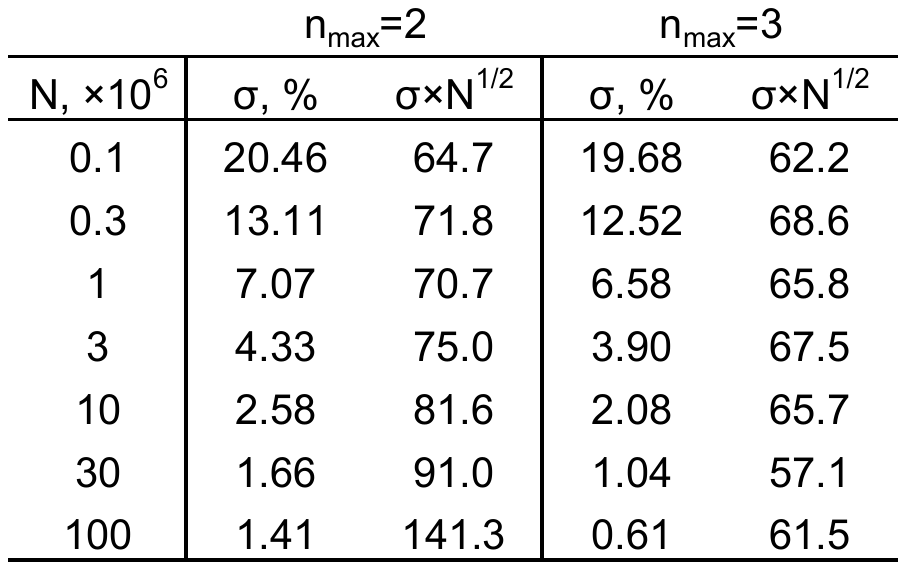}
\end{array}$
\fi
\parbox{0.75\textwidth}{
\caption[Comparison between the accuracies of Monte Carlo simulations and analytical approximation.]
{\emph{Comparison between the accuracies of Monte Carlo simulations and analytical approximation.}}\label{fig:MCvsAN}}
\end{figure}
To assess performance of the proposed approximation, analytically calculated risk contributions have been compared to those calculated by means of Monte Carlo simulation. The comparison has been made using the settings described in Section \ref{sec:portmodel}. In particular, the same valuation function \eqref{eq:defaultval}-\eqref{eq:nodefaultval} has been used in both cases. Analytical risk contributions were computed numerically using the algorithm described in Section \ref{sec:algorithm}, modified by  \eqref{eq:approxulccredit} and \eqref{eq:approxulccredit2}. Monte Carlo risk contributions were estimated using the definition \eqref{eq:ulc} according to
\begin{equation}
\sigma_i^c = \frac{1}{\sigma_p}\cdot\frac{1}{N_{mc}}\sum_{k=1}^{N_{mc}}(v_{ik}-\overline{v}_i)(v_{pk} - \overline{v}_p),\quad \sigma_p = \sum_{k=1}^{N_{mc}}(v_{pk} - \overline{v}_p)^2
\end{equation}
and
\begin{equation}
\overline{v}_i = \frac{1}{N_{mc}}\sum_{k=1}^{N_{mc}}v_{ik}, \quad \overline{v}_p = \frac{1}{N_{mc}}\sum_{k=1}^{N_{mc}}v_{pk},
\end{equation}
where $N_{mc}$ is a number of scenarios and $v_{ik}$ and $v_{pk}$ are loan and portfolio value realizations in the $k$th scenario.

A "real life" portfolio was used for numerical tests.
The portfolio consisted of 8036 loans to 4378 borrowers. Horizon $t_h$ was chosen to be 1 year, 4\% flat risk-free interest rate was used and market price of risk $\lambda$ was assumed to be 0.4. The recovery uncertainty parameter $k$ in \eqref{eq:betameanvar} was set to be the same for all loans and equal to 4.
A wide range of loan risk drivers was covered. Namely, probabilities of default at horizon were in the range from $10^{-5}$ to 0.4, maturities varied between 1 month and 30 years, losses given default $l$ were spread between 0.1 and 0.99. The borrowers represented a wide variety of geographic regions and industries. The borrower-specific coefficients $\{r^2_i\}$, determining dependency on the common factors $\{\eta_k\}$, covered the range from 0.07 to 0.65. The common factor model used in the experiments contained 120 factors $\{\eta_k\}$.

The accuracy of the Monte Carlo based estimate of $\sigma^c$ depends on the number $N$ of scenarios in a simulation. The standard error of the Monte Carlo estimates is proportional to $\frac{1}{\sqrt{N}}$. Several Monte Carlo simulations were performed with different numbers of scenarios. Starting at $10^5$, the number of scenarios was gradually increased up to $10^8$.

The accuracy of the analytically estimated $\sigma^c$, on the other hand, is limited by the number of terms $n_{max}$ in the series expansion \eqref{eq:covarseries} that are taken into account.
The amount of common factors $N_f=120$ in the model imposed a practical limitation on the
cutoff point $n_{max}$ in \eqref{eq:approxulccredit}, since for the $n$th order terms in the series expansion \eqref{eq:covarseries} one has to keep track of $(N_f)^n$ portfolio parameters $P^{(n)}_{k_1\ldots k_n}$. Two values of the cutoff parameter $n_{max}=2$ and $n_{max}=3$ have been considered in the experiments.

Deviations between Monte Carlo and analytical estimates of risk contributions, $\frac{\sigma^c_{mc}-\sigma^c_{an}}{\sigma^c_{an}}$, have been studied for different combinations of $N$ and $n_{max}$. Figure \ref{fig:error} displays the relative differences between the Monte Carlo and analytical estimates of the risk contributions for each loan in the portfolio together with the standard error of this deviation. One can see that when the Monte Carlo noise is suppressed by a high number of scenarios, the $n_{max}=3$ case exhibits better performance in terms of accuracy. The agreement between the two methods in case $n_{max}=3$ and $N=10^8$ is remarkably good.

For both $n_{max}=2$ and $n_{max}=3$ cases the deviations from the $\sigma \sim \frac{1}{\sqrt{N}}$ Monte Carlo error convergence rule have been studied. Figure \ref{fig:MCvsAN} shows the dependency of the $\sigma\cdot\sqrt{N}$\footnote{$\sigma$ is a standard deviation of the relative difference between analytical and simulation-based estimates of risk contributions $\sigma^c$ across the portfolio.} on the number of simulations. For the $n_{max}=3$ case the product $\sigma\cdot\sqrt{N}$ remains constant up to $N=10^8$, which implies that the difference between the Monte Carlo and analytical estimates is mainly due to the noise of the Monte Carlo simulation. Thus, the accuracy of the analytical approximation in case $n_{max}=3$ exceeds the accuracy of Monte Carlo simulation using $10^8$ scenarios.

Computer time spent was 13 seconds for $n_{max}=3$ analytical and 16 hours for $N=10^8$ Monte Carlo calculations.
Overall, the proposed analytical approximation to variance-covariance based risk allocation demonstrated exceptionally good performance both in terms of accuracy and computer time.

\subsection{Accuracy of the approximation}
The high accuracy of the proposed approximation has been demonstrated by benchmarking the analytical estimates against Monte Carlo simulation. However, some explanations are necessary to address possible scepticism regarding the convergence properties of the expansion \eqref{eq:covhermite}.

Indeed, approximating a pairwise correlation by the first three terms of the above mentioned series expansion may lead to a rather rough estimate. Asset correlation $\rho_{ij}$ of two borrowers with high systematic components of asset returns ($r_i$ in \eqref{eq:split}) from the same country/industry cluster (i.e. the same factor loadings $\{\beta_i\}$) may be higher than 0.5, which results in a poor convergence of \eqref{eq:covhermite}. However,
two important mitigating factors contribute to the overall accuracy of the approximation.

First, according to \eqref{eq:ulc}, the risk contribution $\sigma^c_i$ is proportional to a sum of covariances of the facility $i$ with all other facilities in a portfolio. The accuracy in this case depends on the average pairwise asset correlations $\rho_{ij}$ which are quite smaller (compared to 0.5) in realistic portfolios due to country/industry diversification and existence of facilities with low systematic components of asset returns.

Second, contributions $v^{(n)}$ to the higher order ($n \geq 3$) parameters $P^{(n)}_{k_1\ldots k_n}$ in \eqref{eq:portparam} have different signs depending on the valuation function $v_i(\epsilon)$. This \emph{pd diversification} effect is obvious in a default-only case \eqref{eq:defonly} and is caused by oscillating nature of higher order Hermite polynomials. As a consequence, the higher order contributions decrease and overall convergence properties of \eqref{eq:approxulc} are improved.

Both of the above mentioned factors contributing to the accuracy of the proposed approximation are significant if the portfolio consists of large number of loans with different characteristics (country/industry association, probability of default, size of systematic part of asset return, etc). In other words, the more realistic the portfolio, the better the accuracy of the approximation.

\section{Summary}
Variance-covariance based risk allocation method has been proposed. The method was applied to a "real life" credit portfolio model and tested on a "real life" portfolio. It was shown that the proposed analytical approximation is superior both in terms of speed and accuracy to the traditional Monte Carlo simulations.

The proposed routine may be especially appealing for the purpose of risk adjusted loan pricing since it allows to compute accurate and statistical noise-free estimates of capital charges. Once the portfolio-specific parameters \eqref{eq:portparam} are calculated, the risk contribution $\sigma^c$ of a loan can be computed using \eqref{eq:approxulccredit} and the capital charge can be assigned according to \eqref{eq:ec}.

\end{document}